\pdfoutput=1
\documentclass[runningheads]{llncs}
\usepackage{amsfonts}
\usepackage{graphicx}
\usepackage{amsmath}
\usepackage[misc]{ifsym}
\usepackage{dcolumn}
\usepackage{amssymb}
\usepackage{etoolbox}
\usepackage{multirow}
\usepackage{booktabs}
\usepackage{todonotes}
\usepackage{hhline}
\usepackage{changes}
	\definechangesauthor[name={Percusse}, color=orange]{per}
	\setremarkmarkup{(#2)}
\usepackage{array, makecell}
\usepackage{boldline}
\usepackage{colortbl}
\usepackage{hyperref}
\hypersetup{
     colorlinks   = true,
     citecolor    = blue,
     linkcolor = blue,
     anchorcolor = blue,
     filecolor = blue,
     linkbordercolor = blue,
     urlcolor = blue
}
\newcommand\ChangeRT[1]{\noalign{\hrule height #1}}

\begin{document}

\title{Synthesis and Inpainting-Based MR-CT Registration for Image-Guided Thermal Ablation of Liver Tumors}
\titlerunning{Synthesis and Inpainting-Based MR-CT Registration for Thermal Ablation}

\author{Dongming Wei\inst{1,2} \and Sahar Ahmad\inst{2} \and Jiayu Huo\inst{1} \and Wen Peng\inst{3} \and Yunhao Ge\inst{4}  \and Zhong Xue\inst{4} \and Pew-Thian Yap\inst{2} \and Wentao Li\inst{5} \and Dinggang Shen\inst{2}\textsuperscript{(\Letter)} \and Qian Wang\inst{1}\textsuperscript{(\Letter)}}
\authorrunning{D. Wei et al.}

\institute{Institute for Medical Imaging Technology, School of Biomedical Engineering,
Shanghai Jiao Tong University, Shanghai 200030, China\\ \email{wang.qian@sjtu.edu.cn}
\and Department of Radiology and Biomedical Research Imaging Center (BRIC), University of North Carolina
at Chapel Hill, Chapel Hill, NC 27599, USA\\
\email{dgshen@med.unc.edu}
\and North China Electric Power University, Beijing, China
\and Shanghai United Imaging Intelligence Co., Ltd, Shanghai, China
\and Shanghai Cancer Center, Fudan University, Shanghai, China
}
\maketitle              
\begin{abstract}
Thermal ablation is a minimally invasive procedure for treating small or unresectable tumors. Although CT is widely used for guiding ablation procedures, the contrast of tumors against surrounding normal tissues in CT images is often poor, aggravating the difficulty in accurate thermal ablation. In this paper, we propose a fast MR-CT image registration method to overlay a pre-procedural MR (pMR) image onto an intra-procedural CT (iCT) image for guiding the thermal ablation of liver tumors. By first using a Cycle-GAN model with mutual information constraint to generate synthesized CT (sCT) image from the corresponding pMR, pre-procedural MR-CT image registration is carried out through traditional mono-modality CT-CT image registration. At the intra-procedural stage, a partial-convolution-based network is first used to inpaint the probe and its artifacts in the iCT image. Then, an unsupervised registration network is used to efficiently align the pre-procedural CT (pCT) with the inpainted iCT (inpCT) image. The final transformation from pMR to iCT is obtained by combining the two estimated transformations, \textit{i.e.}, (1) from the pMR image space to the pCT image space (through sCT) and (2) from the pCT image space to the iCT image space (through inpCT). Experimental results confirm that the proposed method achieves high registration accuracy with a very fast computational speed.

\end{abstract}
\section{Introduction}
Thermal ablation~\cite{ref_article2} elevates the temperature (55$^{\circ}$--65$^{\circ}$ Celsius) of a focal zone in  the tumor and induces irreversible cell injury and eventually tumor apoptosis and coagulative necrosis. Therefore, accurate targeting of the tumor area is critical for ablating tumor tissues only and leaving the surrounding healthy tissues intact. 

CT imaging is typically used to guide the interventional procedure in thermal ablation, where pre-procedural CT (pCT) is used for planning, and intra-procedural CT (iCT) is captured during the treatment to facilitate safe placement of the ablation probe and accurate targeting of the tumor~\cite{ref_ct}. 
However, CT is relatively poor in tissue contrast (\textit{e.g.,} arteries) and is susceptible to artifacts introduced by the probe during the procedure. Therefore, high-resolution pCT and pre-procedural MR (pMR) images are typically aligned during planning and then registered onto the iCT image for more precise guidance in positioning the probe to the desired region of interest (ROI)~\cite{ref_article3,ref_article4}.
In liver tumor ablation, accuracy and speed of such an alignment are both important as it can compensate for deformations caused by patient positioning and respiratory motion without delay. 

Most volumetric registration algorithms~\cite{ref_article5,ref_article6,ref_article7} are only feasible in the pre-procedural stage as they involve iterative yet time-consuming optimization. Moreover, they do not deal with the probe-induced artifacts. In order to overcome these challenges, we propose a fast image registration framework to align pMR images onto iCT images for guiding thermal ablation of the liver tumor. Meanwhile, our method also eliminates the probe artifacts so that they do not interfere during registration. The proposed registration framework consists of two stages:
\begin{enumerate}
    \item \textbf{Pre-Procedure:} Rigid and deformable registrations between pMR and pCT images. We use mutual-information (MI)-based Cycle-GAN to generate sCT from pMR images to convert the cross-modality registration into a mono-modality problem.
    \item \textbf{Intra-Procedure:} Fast deformable registration of the inpainted iCT (inpCT) image with the pCT image, using an unsupervised registration network (UR-Net). 
\end{enumerate}
Finally, the pMR image is aligned to the iCT image by composing the two transformations estimated in the above two stages.

\begin{figure}[t]
\includegraphics[width=\textwidth]{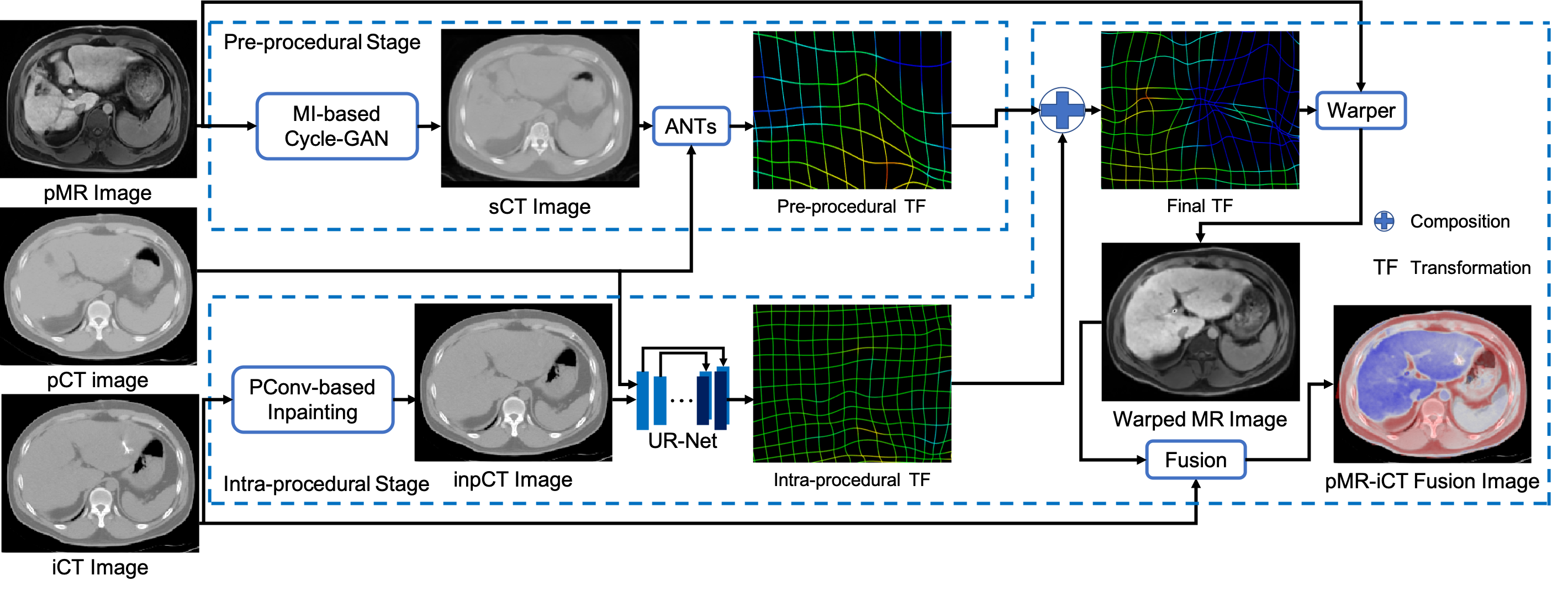}
\caption{The proposed image registration method consists of three deep-learning-based algorithms and the traditional ANTs algorithm.} \label{fig1}
\end{figure}

\section{Methods}
In order to accurately and efficiently register pMR images onto iCT images for guiding thermal ablation of liver tumors, we propose a two-stage registration framework. First, we convert the cross-modality MR-CT image registration into 
\begin{figure}[t]
    \centering
    \includegraphics[width=\textwidth]{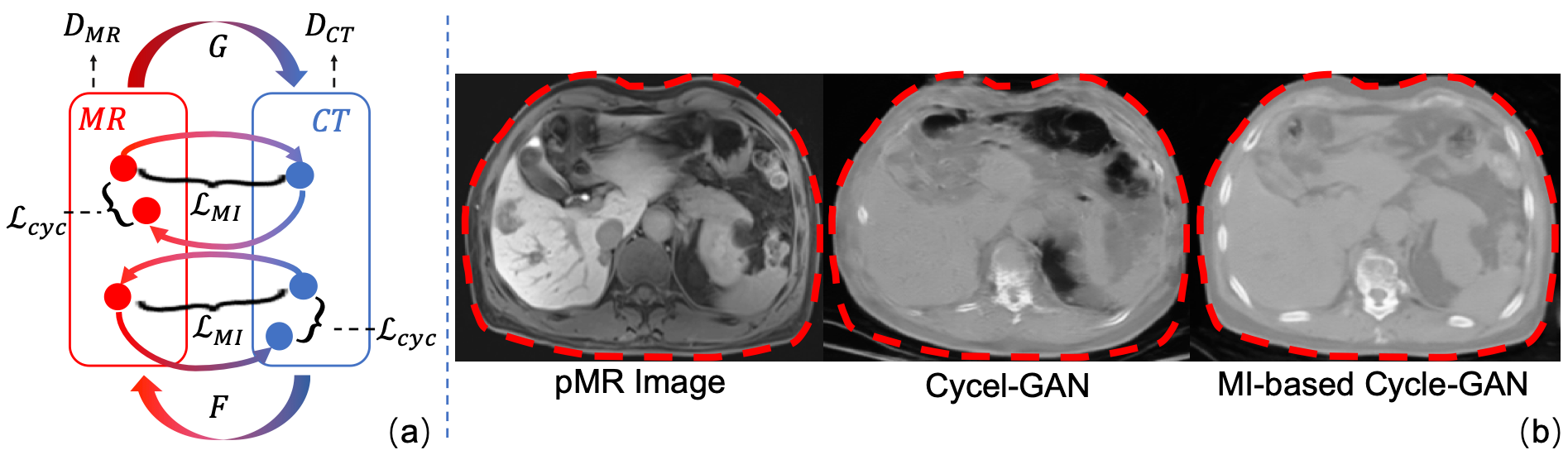}
    \caption{Schematic illustration of MI-based Cycle-GAN. (a) The conventional Cycle-GAN algorithm, and the forward and backward synthesis steps of our method with MI as the explicit structural similarity constraint; (b) exemplar outputs of Cycle-GAN and MI-based Cycle-GAN, where the drifting of the boundaries (highlighted by red contours) can be suppressed by introducing MI-based contraint to Cycle-GAN.}
    \label{fig:MICycle}
\end{figure}
mono-modality registration (CT-CT) by synthesizing sCT images from pMR images through MI-based Cycle-GAN. This mono-modality rigid and deformable registration is performed by using ANTs~\cite{ref_article9}. Then, the trained UR-Net performs deformable registration between pCT and inpCT, which is inpainted by a partial-convolution (PConv)-based network from the iCT image. Thus, the pMR image can be warped onto the iCT space by composing the output transformations of the two registration stages. The pipeline of our method is shown in Fig.~\ref{fig1}.

\subsection{Pre-Procedural MR-CT Registration}
MR-CT registration is challenging due to large appearance differences between the two modalities. Previous works~\cite{ref_articleMRCT} have shown that the cross-modality registration can be converted into a mono-modality registration to achieve better performance. To this end, we synthesize the sCT image from an input pMR image, which then facilitates the subsequent registration between pMR and pCT images.

The MR-to-CT synthesis is completed by using Cycle-GAN with mutual information constraint. Cycle-GAN~\cite{ref_article12}, as one of the state-of-the-art image synthesis algorithms, adopts the adversarial loss given by two discriminators ($D_{\text{MR}}$ and $D_{\text{CT}}$), such that the distribution of the output images of the two generators ($G$ and $F$) is indistinguishable from that of the input images. It also uses the cycle-consistency (shown in Fig.~\ref{fig:MICycle}(a)) to enforce the forward (MR-to-CT) and backward (CT-to-MR) syntheses to be bijective. However, Cycle-GAN fails to enforce structural similarity between the pMR and sCT images, which may lead to uncontrollable drifting of tissue/organ boundaries in the synthesized images (see red dashed contours in Fig.~\ref{fig:MICycle}(b)).

Therefore, in addition to the cycle-consistency loss $\mathcal{L}_{\text{cyc}}$, we propose to introduce the MI loss $\mathcal{L}_{\text{MI}}$ to the generators ($G$ and $F$) to directly enforce the structural similarity between the input and synthesized images (as shown in Fig.~\ref{fig:MICycle}(a)). The MI loss is defined as:
\begin{equation}
\mathcal{L}_{\text{MI}} =\sum\sum p(x,y)\text{log}\frac{p(x,y)}{p(x)p(y)},
\end{equation}
where $p(x)$ and $p(y)$ denote the histograms of $I_{\text{MR}}$ and $G(I_{\text{MR}})$, respectively, and $p(x,y)$ refers to the joint histogram of $I_{\text{MR}}$ and $G(I_{\text{MR}})$. After synthesizing the cross-modality registration (pMR-pCT) is converted into a mono-modality registration problem (sCT-pCT). We then perform a conventional mono-modality registration using ANTs~\cite{ref_article9} to estimate the field that accounts for both rigid and deformable transformations.

\begin{figure}[t]
    \centering
    \includegraphics[width=\textwidth]{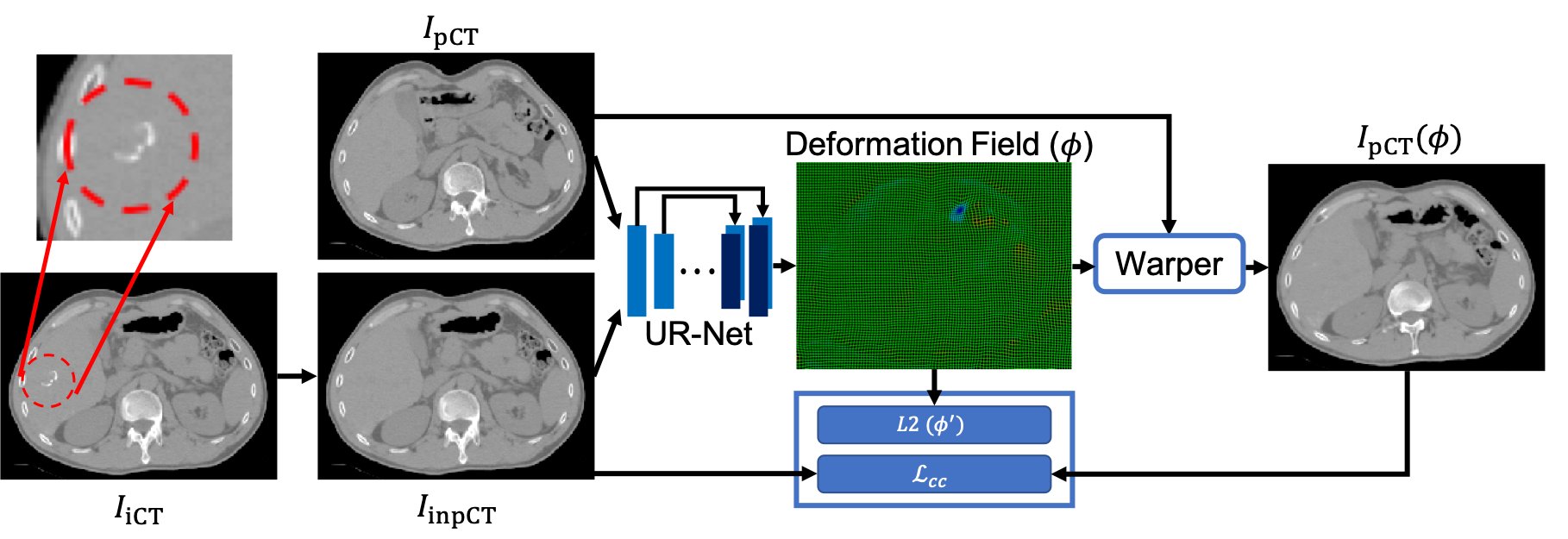}
    \caption{Overview of unsupervised registration. The loss consists of cross-correlation between $I_{\text{pCT}}({\phi})$ and $I_{\text{inpCT}}$ and smoothness of $\phi^{'}$.}
    \label{fig:UR-Net}
\end{figure}

\subsection{Intra-procedural Registration}\label{Section:intra-stage}
Due to the high computational efficiency requirement during intra-procedural stage, we propose to train a UR-Net (as shown in Fig.~\ref{fig:UR-Net}) in an unsupervised manner to perform expeditious deformable registration between pCT and inpCT images. The advantage is its fast speed by using parallel convolution on GPU during network inference. Dalca \textit{et al.} have applied a similar unsupervised registration network to perform brain MR image registration~\cite{ref_article11}, which shows comparable performance with the state-of-the-art optimization-based methods. In our work, the UR-Net is trained by pCT-inpCT image pairs with the loss function defined as:
\begin{equation}
    \mathcal{L} = -\mathrm{S}(I_{\text{inpCT}},I_{\text{pCT}}(\phi)) + \mathrm{Reg}(\phi),
    \label{Equ:loss_of_UR-Net}
\end{equation}
where $I_{\text{inpCT}}$ and $I_{\text{pCT}}$ represent the fixed and moving images, respectively. The loss function comprises of (1) similarity (S) in terms of cross-correlation ($\mathcal{L}_{\text{cc}}$) between the warped moving image ($I_{\text{pCT}}(\phi)$) and the fixed image ($I_{\text{inpCT}}$), and (2) the regularization (Reg) penalty defined in terms of L2-norm of the gradients of the deformation fields ($\text{L}2(\phi^{'})$). The detailed network architecture is shown in \textit{Supplementary} file.

The probe and its artifacts, which hinder the interpretation of the underlying anatomy, may cause inaccurate registration~\cite{ref_article6}. In order to remove the probe and its streak artifacts in the iCT image, we train a PConv-based~\cite{ref_article:Inpainting} 3D U-Net to obtain inpCT image ($I_{\text{inpCT}}$). Specifically, given a rough mask (\textit{e.g.,} a bounding box or a polygon) covering the probe and its artifacts, the inpainting network can reconstruct the underlying tissues and update the mask layer-by-layer until the mask shrinks away. The inpainting is based on 3D convolution (Conv)-based U-Net architecture, while each convolution layer is replaced with partial convolution to ensure that the inpainted contents will not be affected by the probe and its streak artifacts in the mask. The probe and its streak artifacts exist in iCT images but not in pCT images. The mask can be manually drawn in the iCT images by clinicians during the procedure, or prepared in advance as part of the procedure planning.
\section{Experiments and Results}
\subsubsection{Dataset and Pre-processing --} Thirty-nine subjects undergoing liver tumor ablation were included in our experiment. Each subject was scanned with his/her own pMR, pCT and iCT images (see  Table~\ref{tab:data_info} for a summary of the parameters).
Livers were delineated from pMR, pCT and iCT images, respectively, while tumors were delineated from pMR and pCT images but not iCT due to probe artifacts and limited contrast. We used 11 labeled subjects for testing, and the remaining 28 subjects were used for training the three networks. Before normalizing intensities into the range [0,1], the intensities of pCT and iCT images were thresholded in the range [-800,800]. pMR images were also rigidly aligned onto pCT images. Then, all the image sizes were resampled to $256\times256\times128$ with isotropic voxel distances. If not stated otherwise, the same training/testing datasets were used.
\subsubsection{Implementation --} Three networks were trained: (1) pre-procedural stage: MI-based Cycle-GAN network for CT synthesis, and PConv-based U-Net for probe inpainting; (2) intra-procedural stage: UR-Net for deformable registration. All these networks were implemented in Keras and trained on a single NVIDIA Titan X GPU.

\begin{table}[t]
    \centering
    \caption{The in-plane FOV, resolution and scanner used in acquisition of the pMR, pCT and iCT images.}
    \begin{tabular}{c|c|c|c}
    \ChangeRT{1pt}
         &  \textbf{In-plane FOV} & \textbf{Resolution} & \textbf{Scanner}\\ \hline
        \textbf{pMR image} & $320\times260$ & $1.188\times1.188\times3\space{\text{mm}}^{3}$ & Siemens 3.0T Skyra \\
        \textbf{pCT image} & $512\times512$ & $0.7559\times0.7559\times3\space{\text{mm}}^{3}$ & Philips Brilliance 64 CT \\ 
        \textbf{iCT image} & $512\times512$ & $0.7559\times0.7559\times3\space{\text{mm}}^{3}$ & Philips Brilliance 64 CT \\ 
    \ChangeRT{1pt}
    \end{tabular}
    \label{tab:data_info}
\end{table}
\subsubsection{Evaluation Metrics --} We computed the Dice ratio over ROIs and the target registration error (TRE) over several landmarks of livers and tumors, to evaluate the registration accuracy. These two metrics are widely used for evaluation of registration performance, with higher Dice ratio (or lower TRE) characterizing better registration quality.

\begin{figure}[t]
\centering
\includegraphics[width=\textwidth]{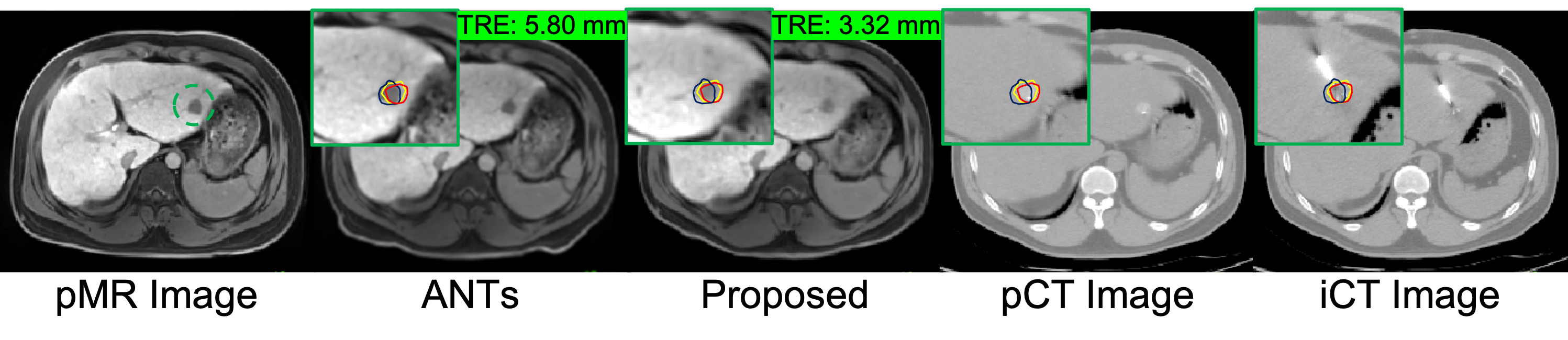}
\caption{Registration results by ANTs and the proposed method. The warped tumor contours by ANTs (blue), our proposed method (yellow) are visualized in zoomed-in view. The tumor contour from manually registered pCT image is used as ground-truth (red). TRE of the tumor center in this case is also reported.} \label{fig3}
\end{figure}

\begin{table}[t]
\caption{Results of pMR-iCT (rigid and deformable), pMR-pCT (rigid and deformable), and pCT-iCT (deformable only) registration.
}\label{tab1}
\centering
\begin{tabular}{cc|c|c|c|c}
\ChangeRT{1pt}
&  & \multicolumn{2}{c|}{\textbf{Rigid}} & \multicolumn{2}{c}{\textbf{Deformable}}\\ \cline{3-6}
 & & \textbf{FLIRT} & \textbf{ANTs} & \textbf{ANTs} & \textbf{Proposed}\\ \hline
 \multicolumn{6}{c}{\textbf{pMR-iCT}}\\ \hline
\multirow{2}{*}{\textbf{Liver}}& \textbf{Dice (\%)} & 48.44$\pm$40.40  & 85.52$\pm$2.92 & {86.59}${\pm}${3.30} & \textbf{86.96}$\boldsymbol{\pm}$\textbf{3.00} \\ \cline{2-6} 

 & {\textbf{TRE (mm)}} & - & 52.56$\pm$9.85 & 5.18$\pm$2.43 & \textbf{4.93$\boldsymbol{\mathrm{\pm}}$2.72}\\\hline 

\multicolumn{6}{c}{\textbf{pMR-pCT}} \\ \hline
\multirow{2}{*}{\textbf{Liver}} & {\textbf{Dice (\%)}} & 48.07$\mathrm{\pm}$40.40 & 87.03$\mathrm{\pm}$2.60 & 89.55$\mathrm{\pm}$2.04 & \textbf{90.59$\boldsymbol{\mathrm{\pm}}$1.73} \\ \cline{2-6}
& {\textbf{TRE (mm)}} & - & 6.63$\pm$2.73 & 5.59$\pm$2.01 & \textbf{4.67$\boldsymbol{\pm}$2.00} \\ \hline
\multirow{2}{*}{\textbf{Tumor}} & {\textbf{Dice (\%)}} & 12.88$\mathrm{\pm}$25.76 & 51.10$\mathrm{\pm}$17.13 & 55.34$\mathrm{\pm}$5.70 & \textbf{62.45$\boldsymbol{\mathrm{\pm}}$4.66} \\ \cline{2-6}
& {\textbf{TRE (mm)}} & - & 6.71$\pm$2.27 & 6.08$\pm$1.40 & \textbf{3.89$\boldsymbol{\pm}$0.99}\\ \hline
\multicolumn{6}{c}{\textbf{pCT-iCT}} \\ \hline

\multirow{2}{*}{\textbf{Liver}}& {\textbf{Dice (\%)}} & - & - & 87.90$\pm$5.25 & \textbf{88.63$\boldsymbol{\pm}$5.53} \\  \cline{2-6}
& {\textbf{TRE (mm)}} & - & - & 5.06$\pm$3.29 & \textbf{4.37$\boldsymbol{\pm}$3.30} \\ \hline
\end{tabular}
\end{table}

\subsection{Registration Results for pMR and iCT}
We conducted pMR-iCT image registration over the test dataset by using FSL FLIRT (rigid), ANTs (rigid and deformable) and the proposed method, respectively. As shown in Fig.~\ref{fig3}, our proposed method achieves better accuracy against ANTs in tumor region and body alignment. Quantitative results using Dice ratio and TRE are presented in Table~\ref{tab1}. It can be observed that our method yields better performance than ANTs. The last line of Table~\ref{tab1} reports the computational time in intra-procedural stage of ANTs and our proposed method, and our proposed method performed the registration in several seconds. Notice that our proposed method in the intra-procedural stage includes the iCT inpainting and the pCT-iCT image registration steps, which were computed on GPU.
\subsection{Pre-procedural Stage}
\subsubsection{Pre-procedural Stage Registration --} The Dice and TRE metrics were evaluated over pMR-pCT pairs and their corresponding sCT-pCT pairs of the testing dataset. As shown in Table~\ref{tab1}, the registration performance of our algorithm is improved over sCT-pCT pair, especially on the target tumor region (more than 7\% Dice and 1.1mm TRE improvement), which proves that our proposed synthesis algorithm can facilitate the cross-modality registration.

\subsubsection{MR-to-CT Synthesis --} We extracted 2240 slices from the transverse planes of 28 unpaired pMR and pCT images and used them as the training dataset. pMR and pCT images of each subject were linearly registered. For testing, we applied the MR-to-CT generator to synthesize CT images from pMR images slice-by-slice and then concatenated the synthesized slices into 3D volumes. As shown in Fig.~\ref{fig:MICycle}(b), our method predicted better CT-like images from pMR images. 

\begin{figure}[t]
\centering
\includegraphics[width=0.85\textwidth]{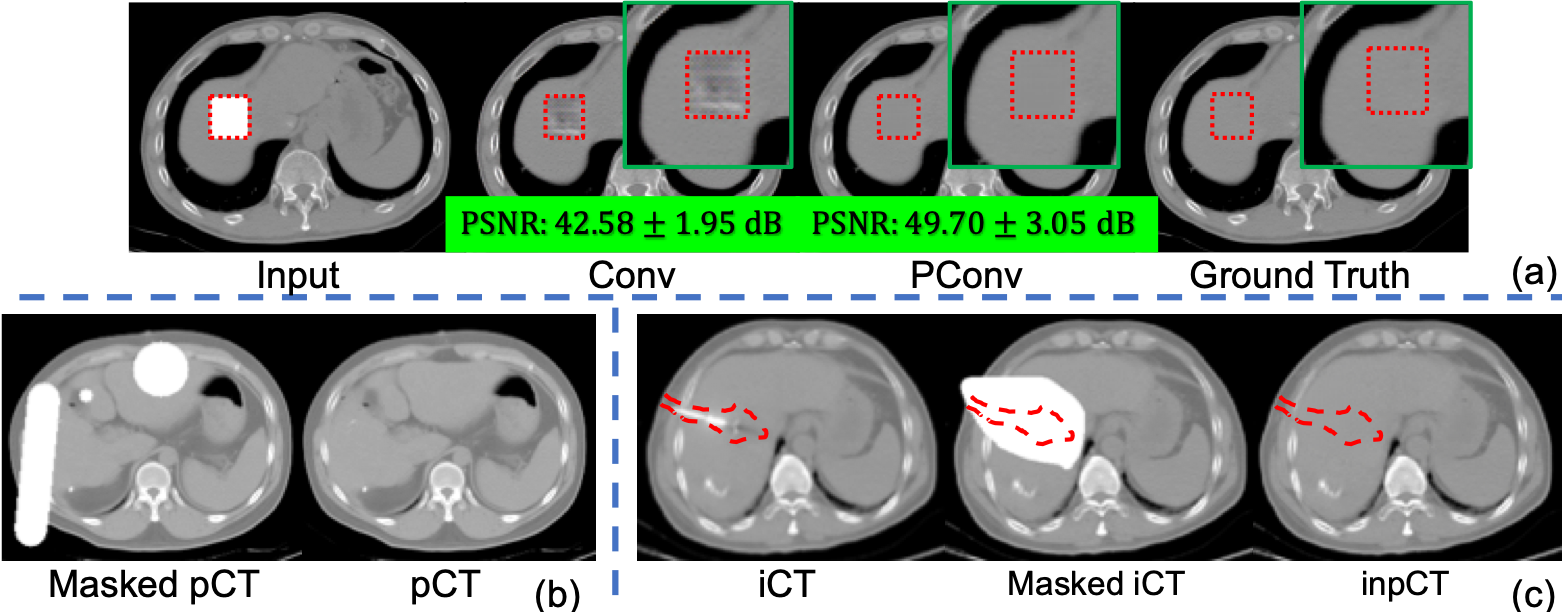}
\caption{
(a) Qualitative and quantitative (mean $\pm$ std PSNR over test subjects) comparison results of Conv-based and PConv-based inpainting network on iCT images; (b) A training pair exemplar; (c) Results of inpainting network in clinical practice.} \label{fig5}
\end{figure}
\subsection{Intra-procedural Stage}
\subsubsection{Intra-procedural Stage Registration --} UR-Net was trained by $28$ rigidly registered pCT-inpCT image pairs in an unsupervised manner, where inpCT images were generated by the trained PConv-base inpainting network. We compared the UR-Net registration performance with ANTs over pCT-inpCT image pairs. The Dice and TRE metrics of liver and the intra-procedural computation time were evaluated (see Table~\ref{tab1}). It can be seen that the UR-Net yielded better performance and was more efficient. The UR-Net method predicted a deformation field in around 3 secs for a $256\times256\times128$ image pair.
\subsubsection{Inpainting --} pCT images were used for training the inpainting network. Each image was augmented 500 times using 2 or 3 random 3D shapes (\textit{i.e.}, 3D balls and bars) with random locations and sizes to imitate manual mask (cf. Fig.~\ref{fig5}(b)). In the testing stage, we compared the inpainting results quantitatively and qualitatively on the iCT images. The results in Fig.~\ref{fig5}(a) shows that PConv-based network obtained around 7 dB PSNR improvement in the masked region than Conv-based network. Particularly, each iCT image was combined with a cubic mask ($40\times40\times20$) in healthy tissue region and input into PConv and Conv-based networks, respectively. The PSNR is computed over the reconstructed region with the ground-truth.
For clinical cases, the results are shown in Fig.~\ref{fig5}(c). The mask was drawn in the pre-procedural images by physicians to cover the tumor and the planned puncture pathway in the tumor-centered transverse plane. Then, an appropriate height was chosen in the intra-procedural stage to ensure that the closed 3D contour can cover the probe and its artifacts. The inference time was around 2 secs for a $256\times256\times128$ subject.
Note that we will further quantitatively evaluate the improvement of tumor registration, and investigate the possibility of reconstructing the tumor directly by the PConv-based inpainting network.
\section{Conclusion}
A learning-based registration framework is proposed to align pre-procedural MR and intra-procedural CT images for image-guided thermal ablation of liver tumor. Experimental results showed that our method can efficiently and effectively overlay pMR onto iCT during ablation with high registration accuracy, compared to the state-of-the-art ANTs algorithm. We also showed that MI-based Cycle-GAN synthesis and unsupervised registration improves the overall performance. 

\subsubsection*{Acknowledgement}
This work was partially supported by the National Key Research and Development Program of China (2018YFC0116400) and STCSM (19QC1400600).

\end{document}